\documentclass[runningheads]{llncs}
\usepackage[T1]{fontenc}
\usepackage{hyperref}
\usepackage{cite}
\usepackage{todonotes}
\usepackage{graphicx}
\usepackage{xspace}
\usepackage{color}
\usepackage{tikz} 
\usepackage{amsmath,amssymb,amsfonts}
\usepackage{tabularx}
\usepackage{array}
\usepackage{algorithm}
\usepackage{multirow}
\usepackage{graphicx}
\usepackage{subfig}
\usepackage{algorithmic}
\usepackage{tcolorbox}
\usepackage{orcidlink}

\usetikzlibrary{shapes.geometric, arrows.meta, positioning}

\tikzstyle{process} = [rectangle, rounded corners, minimum width=3cm, minimum height=1cm, text centered, draw=black, fill=blue!30]
\tikzstyle{arrow} = [thick,->,>=stealth]

\newcommand{\approach}[0]{\textit{POSEIDON}\xspace}
\begin{document}
\title{\approach: Efficient Function Placement at the Edge using Deep Reinforcement Learning}
%
\titlerunning{\approach: Efficient Function Placement at Edge using Deep RL}


\author{Prakhar Jain\inst{1}\orcidlink{0009-0005-4221-2946}
\and
Prakhar Singhal\inst{1}\textsuperscript{\dag}\orcidlink{0009-0009-7268-5508}
\and
Divyansh Pandey\inst{1}\textsuperscript{\dag}\orcidlink{0009-0002-2915-9624}
\and 
Giovanni Quattrocchi\inst{2}\orcidlink{0000-0002-0405-9814}
\and
Karthik Vaidhyanathan\inst{1}\orcidlink{0000-0003-2317-6175}
}

%
\authorrunning{P. Jain et al.}

\institute{
Software Engineering Research Center,  International Institute of Information Technology\\ Hyderabad, India\\
\email{\{prakhar.jain, prakhar.singhal\}@research.iiit.ac.in}, 
\email{divyansh.pandey@students.iiit.ac.in}, \email{karthik.vaidhyanathan@iiit.ac.in}
\and Politecnico di Milano, Dipartimento di Elettronica, Informazione e Bioingegneria \\  Milan, Italy\\ \email{giovanni.quattrocchi@polimi.it}
}

\maketitle

\renewcommand{\thefootnote}{} 
\footnotetext{\textsuperscript{\dag} These authors contributed equally to this work.}
\renewcommand{\thefootnote}{\arabic{footnote}} 
\setcounter{footnote}{0}
\begin{abstract}

Edge computing allows for reduced latency and operational costs compared to centralized cloud systems. In this context, serverless functions are emerging as a lightweight and effective paradigm for managing computational tasks on edge infrastructures.
However, the placement of such functions in constrained edge nodes remains an open challenge. On one hand, it is key to minimize network delays and optimize resource consumption; on the other hand, decisions must be made in a timely manner due to the highly dynamic nature of edge environments.
In this paper, we propose \approach, a solution based on Deep Reinforcement Learning for the efficient placement of functions at the edge. \approach leverages Proximal Policy Optimization (PPO) to place functions across a distributed network of nodes under highly dynamic workloads. A comprehensive empirical evaluation demonstrates that \approach significantly reduces execution time, network delay, and resource consumption compared to state-of-the-art methods.

\end{abstract}

\section{Introduction}
\label{sec:introduction}
Edge computing has emerged as a promising solution to address the limitations of centralized cloud systems, particularly in terms of reducing latency and operational costs. This paradigm shift enables computational tasks to be processed closer to the data source, thereby enhancing the performance and efficiency of applications \cite{DBLP:journals/access/PhamFHPLLHD20}. 
In this decentralized framework, edge nodes, by their nature, are often constrained in terms of resources such as processing power, memory, and storage. Additionally, the workload on these nodes can be highly fluctuating, with varying demands depending on user activities and their dynamic geographical location. This inherent variability necessitates a flexible and efficient approach to managing applications running on edge nodes \cite{zhang2023resource}.

Recently, the serverless paradigm has emerged as a suitable solution for managing applications in edge computing infrastructures \cite{DBLP:journals/internet/RaithND23}. Serverless allows developers to deploy applications as a collection of discrete, self-contained functions that are designed to be lightweight and stateless \cite{wen2023rise}. In the context of edge computing, such functions can be quickly moved across nodes to adapt to the mobility of users and the shifting demands of applications. 

Despite the advantages, dynamically placing serverless functions in edge nodes presents significant challenges \cite{DBLP:conf/pdp/RussoCP23}. Ideally, functions should be placed as close to users as possible to minimize network delays and optimize resource consumption. However, resource-constrained edge  cannot always host all necessary functions and the mobility of users and the heterogeneity of functions in terms of CPU and memory requirements further complicate the problem. Moreover, such a dynamic environment requires timely decisions to cope with fluctuating workloads \cite{nep:24}. In the literature, most of the work exploits combinatorial optimization techniques, such as Integer Programming formulations, to solve the placement problem effectively \cite{DBLP:conf/icsoc/LaiHACHG018, DBLP:journals/tcc/LaiHGCAHY22,DBLP:conf/seams/BaresiHQT22}. While these methods are capable of producing optimal solutions, they are often complex and time-consuming, resulting in slow solution generation. Some approaches have proposed custom heuristics as an alternative \cite{DBLP:journals/tsc/LiuZHXZ23}, but these methods have been demonstrated to produce significantly lower quality placements compared to optimization-based approaches \cite{nep:24}.

In this context, we introduce \approach \footnote{\hyperref[https://github.com/sa4s-serc/poseidon]{https://github.com/sa4s-serc/poseidon}}, a novel solution that utilizes Deep Reinforcement Learning \cite{lillicrap2019continuous} (DRL) to optimize the placement of serverless functions at the edge. \approach specifically leverages Proximal Policy Optimization \cite{schulman2017proximal} (PPO) to distribute functions across a network of nodes, effectively managing highly dynamic workloads. After determining the placement, \approach uses a simplified Mixed Integer Linear Programming (MILP) problem to optimize traffic routing across the different function instances.
The proposed method aims to reduce network delays, improve resource consumption, and produce timely solutions compared to existing state-of-the-art techniques.

The high-dimensional nature of the placement problem and continuous state space render classical RL algorithms impractical. Traditional RL typically relies on discrete state-action spaces, which are unsuitable for complex scenarios such as edge computing. DRL overcomes this limitation by using neural networks to approximate complex mappings from states to actions or action probabilities, enabling efficient and scalable learning in high-dimensional state spaces.
Moreover, being known for its ability to capture intricate interactions between states and actions, DRL is well-suited to handle real-world environments that are complex and dynamic, requiring flexible and adaptive learning methods \cite{lillicrap2019continuous,Mnih2015}. 

We evaluated \approach through a comprehensive comparison with state-of-the-art solutions. Our extensive empirical evaluation demonstrated that \approach is almost 16 times faster than the state-of-the-approach with respect to decision time with almost comparable cost and delay to the state-of-the-art.
%


The rest of the paper is organized as follows. Section~\ref{sec:problem} presents the problem and introduces our solution. Section~\ref{sec:approach} details the DRL solution and the MILP formulation. Section~\ref{sec:results} presents the empirical evaluation of \approach and the comparison against the state-of-the-art. Section~\ref{sec:related_works} describes some relevant work and concludes with Section~\ref{sec:conclusions}.

\section{Problem and Solution Overview}
\label{sec:problem}
In \approach, an edge topology is defined as a graph where $N$ is the set of nodes and the edges are the links between them. Each pair of nodes $i$ and $j$ is characterized by $\delta_{ij}$, representing the network delay between them. $F$ is the set of functions that could be deployed on the edge topology. We assume that users can connect to any of the nodes in $N$ based on their geographical proximity to the closest one. Thus, for each function $f$ and node $i$, the incoming workload for a function $f$ to node $i$ is defined as $w_{f,i}$, representing the number of requests for $f$ arriving at node $i$. Since each node is resource-constrained and cannot host all functions, we assume that each node can route the requests to any other (nearby) node $j$. Thus, the problem \approach tackles is twofold:

\begin{enumerate}
    \item Deciding whether an instance of function $f$ should be placed on node $i$ (\textit{placement});
    \item Deciding how the workload incoming to any node $i$ for function $f$ should be routed to any other node $j$ (\textit{routing policies}).
\end{enumerate}

\subsection{Solution overview}

\begin{figure}[t]
    \centering    \includegraphics[width=\linewidth]{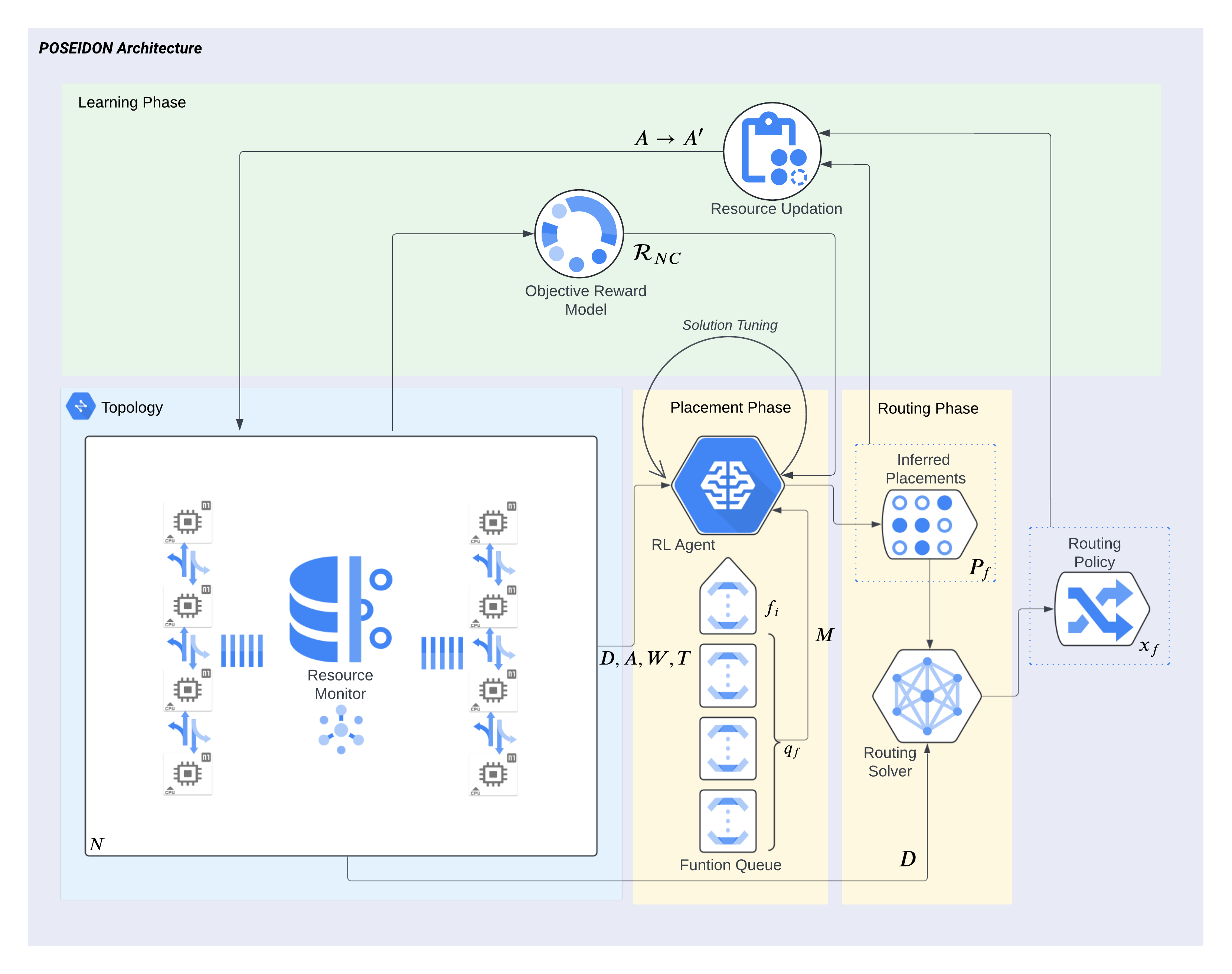}
    \caption{\approach architecture.}
    \label{fig:pos-arch}
\end{figure}

To address these problems, \approach leverages DRL for placement and a MILP formulation for routing policies. The goal of \approach is to minimize both (i) the overall network delay, i.e., the latency while serving function requests, and (ii) the cost of running the function instances. The former goal focuses on placing functions as close to users as possible to minimize routing delays. The latter goal aims to use the minimum number of nodes to serve all the workload with minimal overhead. \approach ensures a balance between these conflicting objectives using a user-defined trade-off.
\approach works in three phases and its architecture is shown in Figure~\ref{fig:pos-arch}. 

The {\em placement phase}, detailed in Section~\ref{subsec:placement}, is dedicated to computing the function placement. To achieve this, \approach organizes the functions $F$ into a queue $q_{F}$, prioritized by their specific criteria such as frequency of requests or resource requirements. Then, a DRL agent considers each function one at a time, starting with the highest priority. For each function, the agent computes a placement vector, a boolean vector $\mathbf{c^f_i}$, which indicates whether a function $f$ should be placed on node $i$.
The agent leverages comprehensive information from the topology, including inter-node delays, available hardware resources, and function parameters such as memory requirements. It also considers the workload of the function instance, reflecting the traffic of function requests at each node. This information is used to infer an optimal function placement that satisfies our objectives.

In the {\em routing phase}, the function placement determined by the agent is used to compute the routing policies, as detailed in Section~\ref{subsec:routing}. In this phase, every node $i$ could host an active function instance and/or route a portion of the requests to another node $j$ with a previously placed $f$ instance. This is done by computing matrix $\mathbf{x^f_{i,j}}$, where each value is a real number between $0$ and $1$ that represents the fraction of the workload incoming to node $i$ for function $f$ that should be routed to node $j$.
To compute the routing policies, \approach utilizes the generated function placements to formulate a MILP problem. The constraints of this problem ensure that all function requests are redirected to nodes with a running instance of the function. The objective is to minimize the network delay of function requests and the cost of running the functions, thereby aligning with the overall solution objective.


In the {\em learning phase}, the state of the topology is updated, and a reward $\mathcal{R}$ is calculated, as detailed in Section~\ref{subsec:comm_state_update_and_reward}. This reward mechanism ensures that the DRL agent learns to place the functions in a manner that minimizes both the total network delay and the operational costs of the function instances.

\section{\approach}
\label{sec:approach}
\begin{table}[h]
\caption{Inputs, Outputs and Reward Variables}
\label{tab:obs_space_components}
    \centering
    \renewcommand{\arraystretch}{1}
    \small
\begin{tabular}{p{3.5cm} >{\centering\arraybackslash}p{1.2cm} >{\centering\arraybackslash} p{5.4cm}}
\hline
\multicolumn{1}{c}{\textbf{Component}} & \multicolumn{1}{c}{\textbf{Symbol}} & \multicolumn{1}{c}{\textbf{Description}} \\
\hline
\textbf{Topology Data}\\
\hline

Inter-Node Delay & $\delta_{i,j}$ & Network delay between node \(i\) and node \(j\) \\
Available node memory & $m_i$  & Available memory on node $i$ \\
Available node cores & $k_i$  & Available CPU cores on node $i$ \\
\hline
\textbf{Function Data}\\
\hline

Function Workload & $w_{f,i}$  & Workload for function $f$ on node $i$ \\[1mm]
Function memory & $m^R_{f}$ & Memory required by function $f$ \\[1mm]
Function cores & $k^R_{f,i}$  & Average CPU cores required by instances function $f$ on node $i$ \\

\hline
\textbf{Monitored Data}\\
\hline
Total Delay & $T$ &  Total network delay \\
Total Cost & $C$ & Total cost of running placed functions \\

\hline
\textbf{Output Data} \\
\hline
Placement vector & $c_{f,i}$ &   Boolean decision variable representing whether function $f$ is placed on node $i$.\\
Routing policy  & $x_{f,i,j}$ &  Real decision variable representing the fraction of $f$ workload incoming into node $i$ to be routed to node $j$ \\
\hline
\end{tabular}
\end{table}

This section details the three phases of \approach. To facilitate the read, we included the most important variables used in our formulation in Table~\ref{tab:obs_space_components}.

\subsection{Function Placement}
\label{subsec:placement}
\approach uses five vectors to represent the state of the topology, $D$, $A$, $W$, $M$, $T$. $D$ is the delay vector where each element is the inter-node delay $\delta_{i,j}$, $A$ is the available resources, $W$ is the workload of the current function $f$, $M$ encodes the parameters of the current function and $T$ denotes the cumulative delay of the topology. The delay vector $D$ can be represented as: 
$$
D = 
\begin{bmatrix}
\delta_{1,1} & \dots & \delta_{1,n} &
     \delta_{2,1} & \dots & \delta_{2,n} & \dots\dots &     \delta_{n,1} & \dots & \delta_{n,n}
\end{bmatrix}$$
with $\delta_{i,i}=0$ (i.e, local communications do not have delays) and $\delta_{i,j} =  \delta_{j,i}$ (i.e., delay is symmetric) for any node $i$ and $j$.  

The available resources in the topology $A$, consist of the available memory $m_i$ and available CPU cores $k_i$ of each node $i$. This variable captures the available hardware resources at each node and is updated after placing each function instance to account for the resources consumed by the placed functions. Formally,

$$
A = \begin{bmatrix} k_1& m_{1} & k_2& m_{2} & \dots & k_n& m_{n} \end{bmatrix}
$$
The workload of the current function $f$, which is denoted by $W$, is composed of the function workload at each node $i$ denoted by $w_{f,i}$. Such data helps determine whether a function needs to be placed on a specific node based on the amount of function requests being received by it. The function parameters are passed into the state vector in the form of a vector $M$ consisting of $m^R_{f}$, $m^R_{\mu}$, $m^R_{\sigma}$ where the latter two represent the mean and standard deviation of the resource requirements required by the remaining functions in $q_{F}$. These guide the placement decisions while keeping track of the resource requirements of the remaining functions.

\smallskip
$$
M = \begin{bmatrix} m^R_{f} & m^R_{\mu} & m^R_{\sigma} \end{bmatrix}
$$

\smallskip
The total network delay $T$ of the topology initially set to $0$ is continuously monitored and updated after placing each function $f$ as follows:

$$
    T = T + \sum_{i}^{N} \sum_{j}^{N} x_{f,i,j}*w_{f,_i}*\delta_{i,j}
$$

where ${x}_{f,i,j}$ is the routing variable for routing requests of function $f$ from node $i$ to $j$ and is computed after the second phase, as described in Section~\ref{subsec:routing}

Feeding the cumulative total delay to the agent helps it learn to minimize the delay effectively based on the existing network delay of the system.

The DRL agent implemented in \approach system employs a policy gradient method known as Proximal Policy Optimization (PPO) to learn an optimal policy $\pi$. This policy maps a given state $s \in \mathcal{S}$ of the topology (i.e., the execution environment) to an action $c \in \mathcal{A}$, where $\mathcal{A} \in \{0, 1\}^N$ represents the action space of the agent. This action space comprises of a set of boolean values for each node, indicating whether a function instance should be placed on the corresponding node or not. The state $s$ exists within a continuous state space and encapsulates environmental information that affect the placement of the function and is formally defined as a feature vector as follows:

$$
    s = \begin{bmatrix}
    D & A & W & M & T \\
    \end{bmatrix}^{\top}
$$

Since $s$ is high dimensional and in a continuous state space it is difficult to model the placement problem with RL methods such as Q-Learning which requires tabulation and hence impractical to store and update the table, the need for DRL arises. The agent consists of a neural network that takes $\mathcal{S}$ as input and estimates a policy $\pi^{*}$ which infers actions $\mathcal{A^{*}}$. These actions are used to compute the reward $\mathcal{R}$ for the agent after a suitable routing policy is computed, as mentioned in Section~\ref{subsec:comm_state_update_and_reward}.

\subsection{Routing policies}
\label{subsec:routing}

Function placement is not sufficient to minimize the total network delay of the system because a node with the running instance of the function may not be able to handle the heavy workload expected in edges networks. In such cases, the need to route a certain fraction of function requests to other available nodes arises which would prevent overloading the node with user requests. To handle the routing of function requests, the objectives are formulated as a Mixed Integer Programming problem where the constraints are defined as follows:

\begin{equation}
    P_f = \{ i \mid c_{f,i} = 1, \, i \in N \} \quad  \forall f \in F
    \label{eq:chosen_nodes} 
\end{equation}
where $P_f$ is the set of chosen nodes for placing a function. Thus, the following constraint ensures that all the $f$ workload  from node $i$ are routed only to the nodes that belong to $P_f$.

\begin{equation}
\sum_{j}^{P_f} x_{f,i,j} = 1 \quad \forall i \in N, \forall f \in F 
\label{eq:chosen_nodes_sum}
\end{equation}

On the contrary, the following equation states that no fraction of $f$ workload should be routed from any node $i$ to a node $j$ that does not belong to $P_f$.

\begin{equation}
 x_{f,i,j} = 0 \quad  \forall i \in N, \forall f \in F, \forall j \notin P_f
\label{eq:chosen_nodes_zero}
\end{equation}

For each node $j$ that host a function $f$, the required CPU cores to handle the total workload forwarded to the node should be lower than the available cores on that node. This is given by:

\begin{equation}
\sum_{i}^{N} x_{f,i,j} * w_{f,i} * k^{R}_{f,j} \leq k_j \quad \forall f \in F, \forall j \in P_f 
\label{eq:core_constraint}
\end{equation}

The objective of the problem is to minimize the network delay of the placed function. The network delay $T_{f}$ for a specific function $f$ is defined as:
\begin{equation}
    T_{f} = \sum_{i}^{N}\sum_{j}^{N} x_{f,i,j} * w_{f,i} * \delta_{i,j}
\end{equation}

\subsection{State Update and Solution Tuning}
\label{subsec:comm_state_update_and_reward}


\begin{algorithm}[t]
\caption{Update state after placing function $f$}\label{alg:updateState}
\begin{algorithmic}[1]
\REQUIRE $A$, $M$, $q_F$, $P_f$
\FOR{each $i \in P_f$}
    \STATE $m_{i} \gets m_{i} - m^R_{f}$
    \STATE $k_{i} \gets k_i - \sum_{j}^{N} w_{f,j} * x_{f,j,i} * k^R_{f,i}$
\ENDFOR 
\STATE $pop(q_F)$
\STATE $f = \textbf{peek}(q_F)$
\end{algorithmic}
\end{algorithm}

Based on the updated state of the community, the reward $\mathcal{R}$ is calculated to facilitate the parameter updates and solution tuning (Algorithm~\ref{alg:reward}). Solution tuning involves training on the current state $s$ of the system to improve the DRL agent
. The reward model for the environment is designed to ensure that the agent comprehends the objectives of the problem and refines the policy $\pi^*$. The agent receives different types of rewards based on its decisions. The network delay and the system's operational cost are incorporated into a reward term, formulated to be minimized by the agent to improve it's performance. Given the conflicting nature of minimizing network delay and system cost, a user-defined trade-off parameter $\alpha$ is introduced as coefficient of trade off with value in range $(0, 1)$ where increasing $\alpha$ corresponds to increasing weightage of cost minimization objective and decreasing $\alpha$ corresponds to increasing weightage of delay minimization objective. Mathematically, the reward term is defined as:

\begin{algorithm}[t]
\caption{Episodic reward calculation after placing function $f$}
\label{alg:reward}
\begin{algorithmic}[1]
\STATE \textbf{Input:} $T$, $C$, $T_{\text{min}}$ (minimum observed total network delay),  $T_{\text{max}}$ (maximum observed total network delay), $C_{\text{min}}$ (minimum observed cost) $C_{\text{max}}$ (maximum observed cost) 
\STATE \textbf{Output:} $T^{'}$ (normalized total network delay), $C^{'}$ (normalized cost),  $\mathcal{R}$ (reward)
\STATE Calculate $T^{'}$ and $C^{'}$ using:
\[
T^{'} = 2 \left( \frac{T - T_{\text{min}}}{T_{\text{max}} - T_{\text{min}}} \right) - 1 \hspace{13pt}        
C^{'} = 2 \left( \frac{C - C_{\text{min}}}{C_{\text{max}} - C_{\text{min}}} \right) - 1
\],

\IF{size($P_f$) = 0 or $m_i$ < 0 or $k_i$ < 0 or $RoutingSolver.STATUS == INFEASIBLE$}
    \STATE $\mathcal{R}$ = $\mathcal{R}_{penalty}$
\ELSE
    \STATE $\mathcal{R} =  \mathcal{R}_{NC}$
\ENDIF
\STATE Update $T_{\text{min}}$, $T_{\text{max}}$, $C_{\text{min}}$, and $C_{\text{max}}$.
\IF{$q_F$ is empty}
    \STATE \textbf{Return Reward and end episode}
\ELSE
    \STATE \textbf{Return Reward and repeat the process.}
\ENDIF

\end{algorithmic}
\end{algorithm}
\begin{equation}
    \mathcal{R}_{NC} = - (\alpha C' + (1 - \alpha) T')
\end{equation}

where $T'$ and $C'$ are normalized values of the total network delay and the system's cost, respectively. Normalization is employed to linearly scale these terms between [-1,1], ensuring equal upper and lower bounds for both objectives. This solves the scaling problem for different units, thereby maintaining a balanced consideration of both network delay and cost (line 3 of Algorithm~\ref{alg:reward}).
The maximum processing times, denoted as $T_{max}$ is computed based on the edge-case scenario. This scenario assumes a complete cyclic routing, wherein each request must traverse through all nodes before being processed, whereas $T_{min}$ is set to zero reflecting case where all requests can directly be served without need of routing any to other nodes.

$\mathcal{R_{NC}}$ is not sufficient to guide the agent for the objectives of \approach. The agent may attempt to exploit the system by not placing the functions on any nodes, or it may make invalid decisions, such as placing functions on nodes with insufficient compute power or generating placements that result in infeasible mixed-integer programming (MIP) solutions during the routing phase. To discourage such undesirable behavior, the agent receives a a negative reward for each invalid placement decision made by the agent which involves violating compute resource, violating memory resources and violating routing constraint represented by $R_{Penalty}$ (line 5 of Algorithm~\ref{alg:reward}), strictly guiding it away from these solutions.

$$
R_{Penalty} = N(\Omega_i)
$$
where $\Omega_i $ denotes an invalid placement decision.



\section{Evaluation}
\label{sec:results}
The objective of the experiments is to evaluate the effectiveness and efficiency of our approach by answering the following questions:

\smallskip
\noindent \textbf{RQ1.} How does \approach compare to state-of-the-art solutions in terms of delay and cost?

\smallskip
\noindent \textbf{RQ2.} How does \approach perform with respect to decision time compared to state-of-the-art solutions?

\smallskip
\noindent \textbf{RQ3.} How does solution tuning in \approach mitigate invalid placements?

\subsection{Experimental Setup}
\label{subsec:exp_setup}
\smallskip
\noindent \textbf{Execution Environment:}
We implemented a simulated edge environment using the Gymnasium by providing the specifications of the nodes and functions as inputs to the simulation. Function requests were sampled using the Cabspotting ~\cite{Cabspotting} dataset, wherein the node delays between the nodes were predefined. The DRL agent was configured to learn within the Gymnasium environment~\footnote{https://gymnasium.farama.org}, employing the Proximal Policy Optimization (PPO) algorithm, as implemented by the Stable-Baselines3 library~\footnote{https://github.com/hill-a/stable-baselines}. 
The routing solver was integrated into the system using the Linear Solver provided by Google's OR-Tools. The \approach was trained for 50 workload distributions at different timesteps and then evaluated on 150 different workload samples. 
Our simulation was run on an Ubuntu 23.04 machine with 16 GB RAM, powered by a 4.6 GHz 12th Gen Intel i7 processor, and an Nvidia RTX 3060 GPU with 6 GB VRAM.



%

%
\smallskip
\noindent \textbf{Experiment Candidates:}
We evaluated \approach \footnote{\hyperref[https://github.com/sa4s-serc/poseidon]{https://github.com/sa4s-serc/poseidon}} by performing four simulations this was done by placing different number of functions on 5 nodes equipped with \texttt{50, 50, 50, 25, 100} cores and memory capacities of \texttt{100, 100, 200, 50, 500 GB}. The first two simulations, with $\alpha = 0$ and $\alpha = 0.5$, involved placing a \textit{small payload} of 4 functions with memory requirements of \texttt{50, 10, 10, 10 GB} whereas the third and the fourth simulation with $\alpha = 0$ and $\alpha = 0.5$ respectively involved placing a \textit{large payload} of 10 functions with memory requirements of \texttt{10 GB} each on the same set of nodes as described earlier. For each simulation, the agent was trained using 50 workloads for 2000 timesteps, with each timestep lasting 100 seconds. The data used for training came from the \textit{Cabspotting}~\cite{Cabspotting} dataset. The purpose of these simulations was to assess \approach's efficiency in not only minimizing the cost and delay but the decision time as well.


\smallskip
\noindent \textit{1. CR-EUA} : Criticality-Awareness Edge User Allocation~\cite{liu2021criticality}: an allocation strategy tailored for safety-critical, low-latency applications where meeting strict performance and reliability requirements is paramount. This approach aims to maximize the number of requests processed at the highest level of criticality by intelligently assigning edge resources to the most critical tasks. By prioritizing requests based on their criticality levels, CR-EUA ensures that applications requiring immediate attention
receive the necessary computational resources to function optimally. 

\smallskip
\noindent \textit{2. VSVBP} (Variable Sized Vector Bin) \cite{lai2018optimalvsvbp}: a placement and routing approach designed to optimize resource utilization in edge computing networks. It works by maximizing the number of allocated service requests while minimizing the number of active edge nodes, thereby reducing operational costs and energy consumption. The method ensures that the response times of deployed services stay within acceptable limits by efficiently distributing workloads across the available resources. VSVBP models the resource allocation problem as a variable-sized bin packing issue, 
which makes it ideal for scenarios where both performance and cost-effectiveness are critical

\smallskip
\noindent \textit{3. NEPTUNE} \cite{DBLP:conf/seams/BaresiHQT22,nep:24}: a solution designed for optimal placement of serverless functions on edge nodes using MIP with the goal of minimizing both network latency and resource utilization. NEPTUNE optimizes the placement of functions by considering factors like network delay and minimizing the number of active nodes to reduce operational costs. After the initial placement, it uses a second optimization step to reduce service disruptions. Further, 
NEPTUNE also generates routing policies that direct traffic to the appropriate nodes, balancing the workload and minimizing inter-node delays, while ensuring functions are only terminated after they complete their current tasks.


\smallskip
\noindent \textbf{Evaluation Metrics}
To measure the effectiveness and efficiency of the approach, we use three different metrics: \textit{i)} Total Delay: this represents the sum of the delay for each of the function requests in the topology, \textit{ii)} Cost: this accounts for the total cost of running the function instances on the nodes, \textit{iii)} Decision time: this metric evaluates how quickly the placement and routing of functions are determined, reflecting the computational efficiency of the approach.

\subsection{Delay and cost analysis (RQ1)} 
\label{subsec:rq1_delay_and_cost}
\begin{figure}[t]
    \centering
    \subfloat[Box plot depicting the delay while using each of the different approaches]{
        \includegraphics[width=0.45\textwidth]{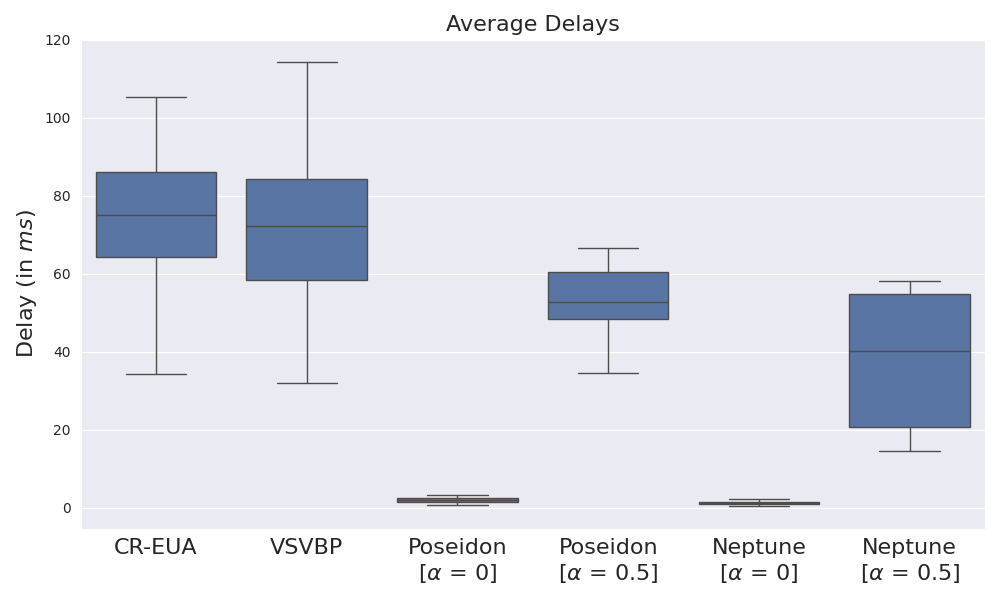}
        \label{fig:light:boxplot_delay}
    }%
    \hfill
    \subfloat[Box plot depicting the cost while using each of the different approaches]{
        \includegraphics[width=0.45\textwidth]{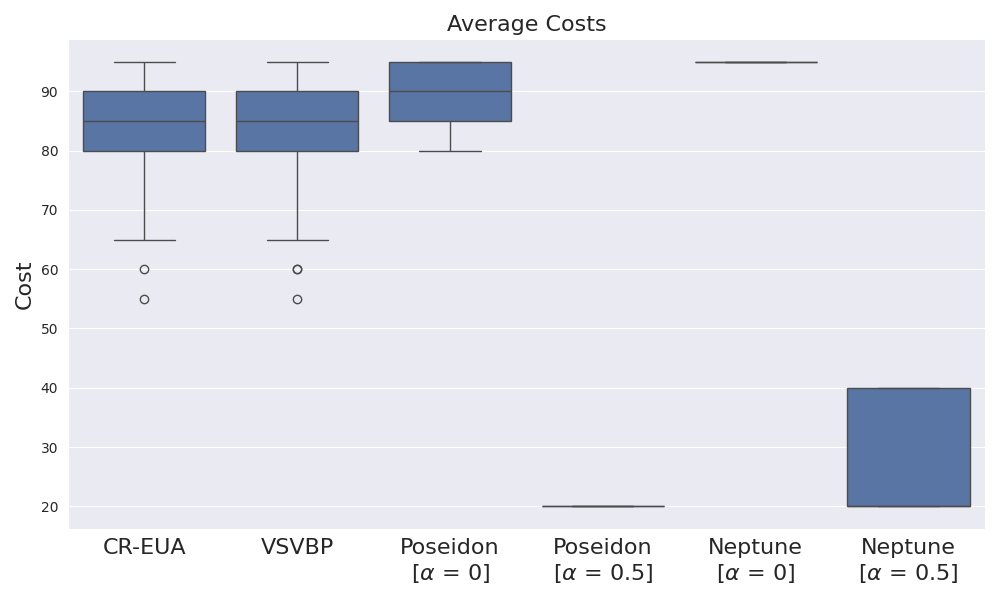}
        \label{fig:light:boxplot_cost}
    }%
    
    \subfloat[Box plot depicting the average decision time while using each of the different approaches]{
        \includegraphics[width=0.45\textwidth]{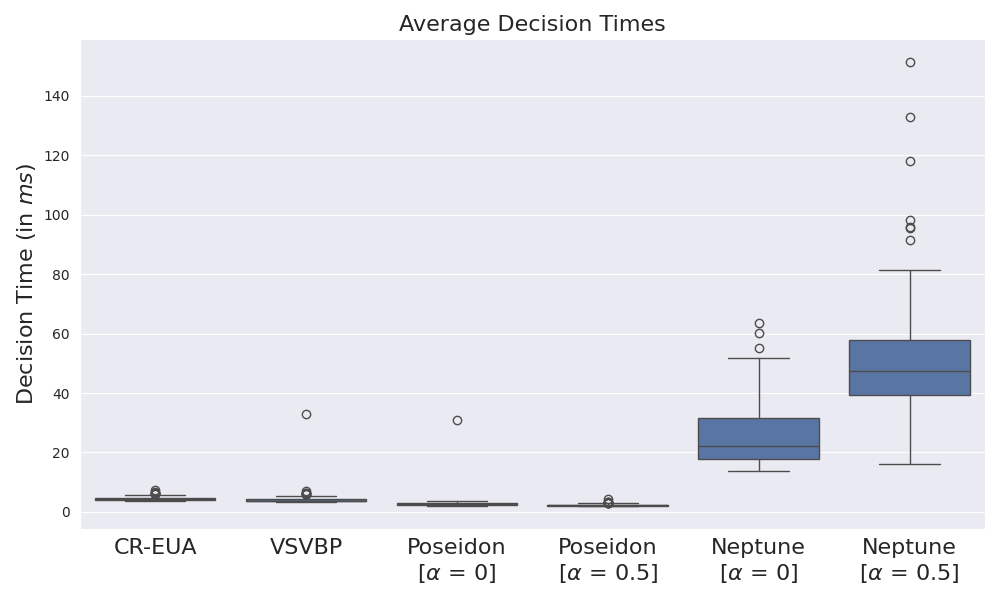}
        \label{fig:light:boxplot_decision_time}
    }
    \caption{Effectiveness of the approaches with respect to various metrics for the \textit{small payload}}
    \label{fig:light:eff_all}
\end{figure}

\begin{figure}[t]
    \centering
    \subfloat[Box plot depicting the delay while using each of the different approaches]{
        \includegraphics[width=0.45\textwidth]{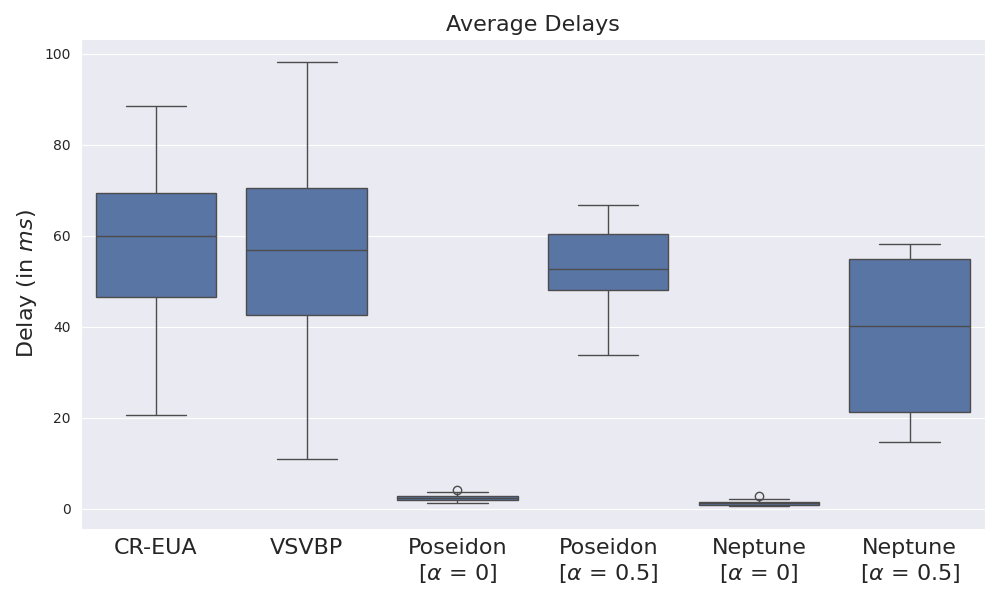}
        \label{fig:heavy:boxplot_delay}
    }%
    \hfill
    \subfloat[Box plot depicting the cost while using each of the different approaches]{
        \includegraphics[width=0.45\textwidth]{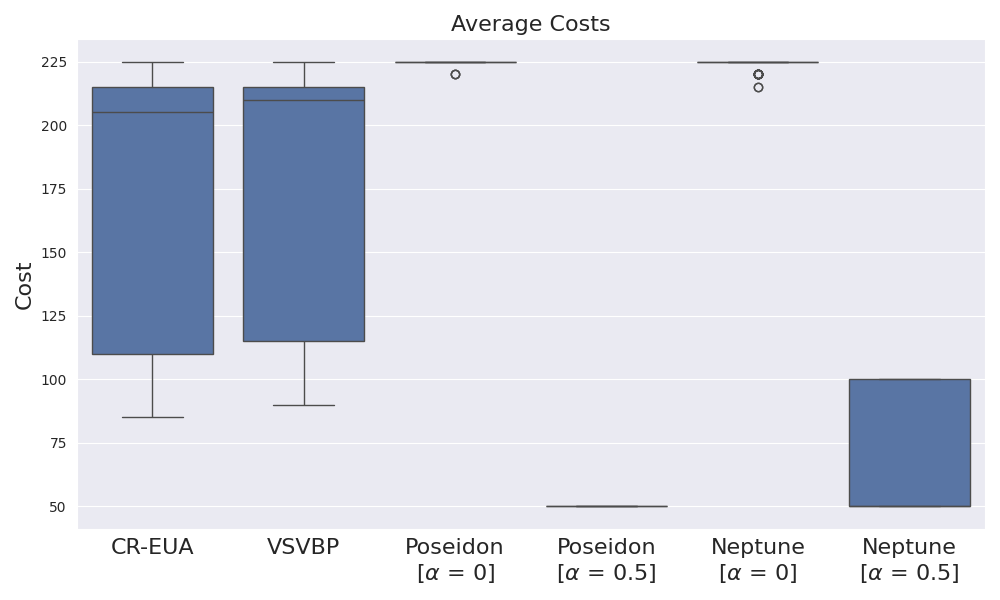}
        \label{fig:heavy:boxplot_cost}
    }%

    \subfloat[Box plot depicting the average decision time while using each of the different approaches]{
        \includegraphics[width=0.45\textwidth]{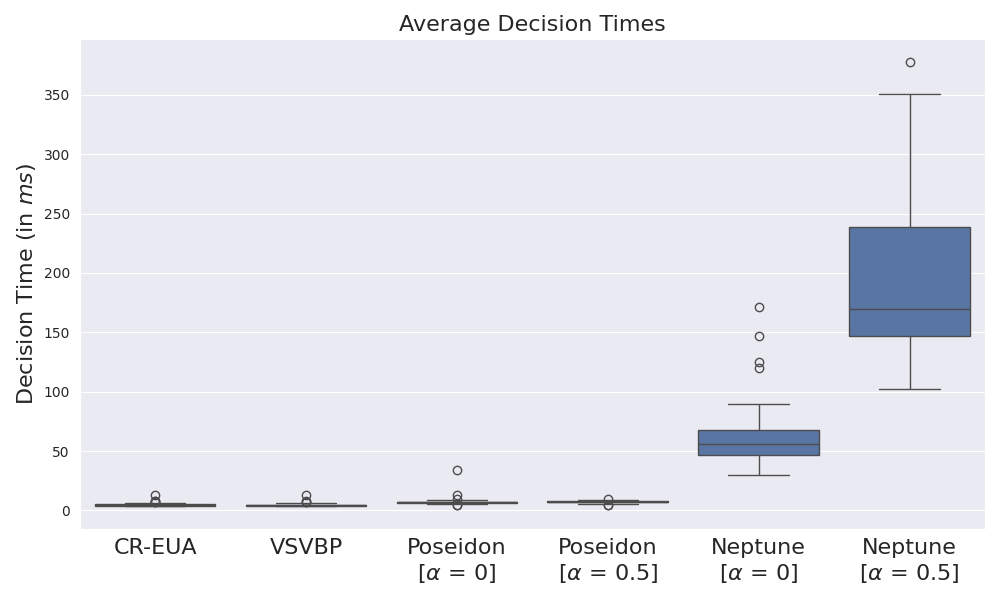}
        \label{fig:heavy:boxplot_decision_time}
    }
    \caption{Effectiveness of the approaches with respect to various metrics for the \textit{large payload}}
    \label{fig:heavy:eff_all}
\end{figure}

To compare the effectiveness of \approach for cost and delay, we evaluated \approach with four different simulations as mentioned in Section~\ref{subsec:exp_setup}. We can see from the Tables~\ref{tab:light:comparison_of_poseidon_and_neptune} and~\ref{tab:heavy:comparison_of_poseidon_and_neptune} and Figures~\ref{fig:light:eff_all} and~\ref{fig:heavy:eff_all} that when $\alpha = 0$ in case of both \textit{large} and \textit{small payload} the average delay is very low but the average cost is high for both \approach and NEPTUNE. This is expected because for $\alpha = 0$ both \approach and NEPTUNE try to minimize the delay and the cost is not considered a critical factor leading to the high average cost. Also both \approach and NEPTUNE outperform the other approaches when it comes to delay for similar average costs. VSVBP and CR-EUA exhibit higher delays because they prioritize maximizing resource utilization and processing the most critical tasks. This focus on handling critical requests often leads to resource overloading, increasing processing times. On increasing $\alpha$ from $0$ to $0.5$, averaging over \textit{small} and \textit{large} payloads, the average costs are much lower, dropping by 77.74 \% for \approach and 71.52\% for NEPTUNE, the average delays increases by 24.9 times for \approach and 33.69 times for NEPTUNE, as both approaches try to find a balance between the delay and the cost. Albeit overall, \approach has a lower cost and higher delay than NEPTUNE. Even though \approach is not as optimal as NEPTUNE but is comparable to NEPTUNE. CR-EUA and VSVBP exhibit higher costs than \approach because they prioritize processing the maximum number of requests, which requires allocating additional resources. Moreover, the stability of \approach can be attributed to its use of PPO, which is known for its robustness and ability to produce stable policies. The training of the DRL agent ensures that that \approach learns to handle varying conditions in a controlled manner.

\subsection{Decision time analysis (RQ2)}
\label{subsec:rq2_decision_time}

To evaluate the efficiency of \approach with respect to decision time compared to other approaches, we ran four different simulations as mentioned in~\ref{subsec:exp_setup} we can see from the Tables~\ref{tab:light:comparison_of_poseidon_and_neptune} and~\ref{tab:heavy:comparison_of_poseidon_and_neptune} and Figures~\ref{fig:light:boxplot_decision_time} and~\ref{fig:heavy:boxplot_decision_time}
\approach demonstrates consistent decision times for $\alpha = 0$ and $\alpha = 0.5$ for both the \textit{small payload} and \textit{large payload} whereas for NEPTUNE the decision time is higher. This can be attributed to the fact that \approach divides the solution into two parts: i)~the DRL agent solves the placement problem and has a very small number of parameters, which enables faster decision-making with regards to placements, and ii) the routing solver which solves the routing problem uses MILP but has fewer constraints than NEPTUNE making \approach 16.43 times faster. 
This also further highlights the scalability challenges incurred by NEPTUNE due to the computational complexity of solving the MILP problem with too many constraints. The other approaches, CR-EUA and VSVBP show comparable decision times to \approach but mostly give sub-optimal solutions with respect to cost and delay.

\renewcommand{\arraystretch}{1.5}
\begin{table}[t]
\scriptsize
\centering
\caption{Comparison of \approach with Other Approaches for \textit{small payload} $\alpha = 0$ and $\alpha = 0.5$}
\begin{tabular}{|l|c|c|c|c|c|c|}
\hline
\multirow{2}{*}{\textbf{Metric}} & \multicolumn{2}{c|}{\textbf{Poseidon}} & \multicolumn{2}{c|}{\textbf{Neptune}} & \multicolumn{1}{c|}{\centering \textbf{CR-EU}A} & \multicolumn{1}{c|}{\centering \textbf{VSVBP}} \\
\cline{2-5}
 & $\alpha = 0$ & $\alpha = 0.5$ & $\alpha = 0$ & $\alpha = 0.5$ & & \\
\hline
Average Delay (in $ms$) & 1.9420 & 53.1460 & 1.1060 & 38.7830 & 74.4574 & 71.6157 \\
Average Cost & 90.7000 & 20.2000 & 95.9500 & 26.4000 & 85.8500 & 85.1000 \\
Average Decision Time (in $ms$) & $3.1$  & $2.4$ & $26$ & $52.1$ & $4.5$ & $4.5$ \\
\hline
\end{tabular}
\label{tab:light:comparison_of_poseidon_and_neptune}
\end{table}

\renewcommand{\arraystretch}{1.5}
\begin{table}[t]
\scriptsize
\centering
\caption{Comparison of \approach with Other Approaches for \textit{large payload} $\alpha = 0$ and $\alpha = 0.5$}
\begin{tabular}{|l|c|c|c|c|c|c|}
\hline
\multirow{2}{*}{\textbf{Metric}} & \multicolumn{2}{c|}{\textbf{Poseidon}} & \multicolumn{2}{c|}{\textbf{Neptune}} & \multicolumn{1}{c|}{\centering \textbf{CR-EUA}} & \multicolumn{1}{c|}{\centering \textbf{VSVBP}} \\
\cline{2-5}
 & $\alpha = 0$ & $\alpha = 0.5$ & $\alpha = 0$ & $\alpha = 0.5$ & & \\
\hline
Average Delay (in $ms$) & 2.3156 & 51.8161 & 1.1744 & 37.6083 & 57.0778 & 54.1870 \\
Average Cost & 227.3333 & 50.5556 & 226.5000 & 66.6667 & 167.7222 & 178.1667 \\
Average Decision Time (in $ms$) & $7.3$ & $7.3$ & $61.7$ & $190.6$ & $4.7$ & $4.6$ \\
\hline
\end{tabular}
\label{tab:heavy:comparison_of_poseidon_and_neptune}
\end{table}
\subsection{Impact of solution tuning (RQ3)}
\begin{figure}[t]
    \centering
    \includegraphics[width=0.75\linewidth]{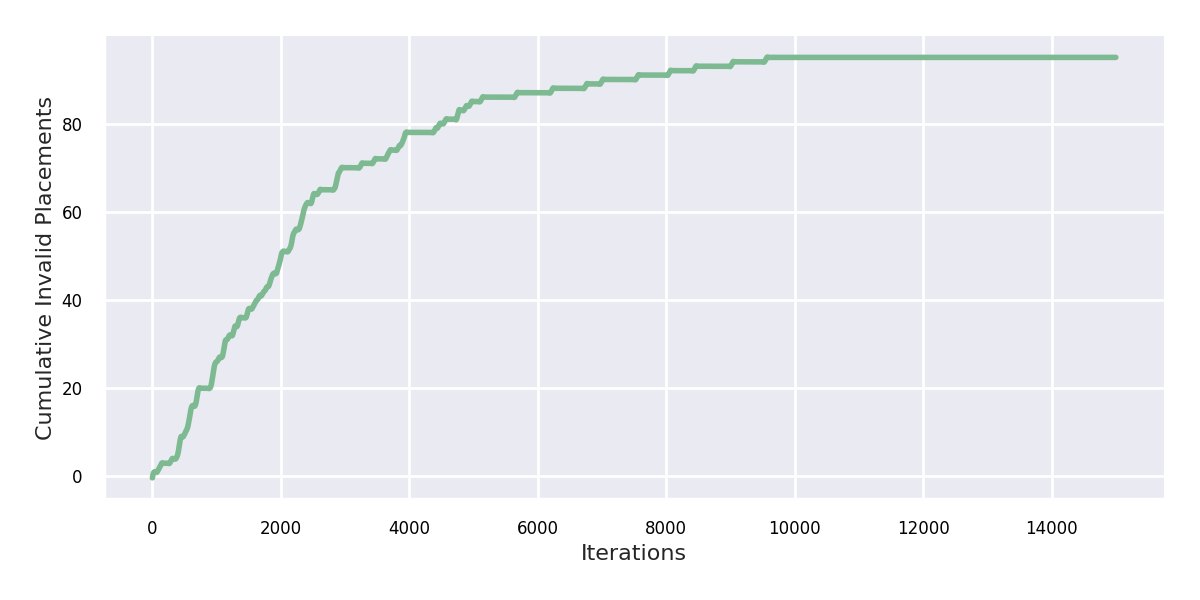}
    \caption{Cumulative count of invalid placements versus iterations of solution tuning for \textit{small payload} (\(\alpha=0\))}
    \label{fig:invalidplacement}
\end{figure}
Figure \ref{fig:invalidplacement} shows the cumulative number of invalid placements performed by the agent against the iterations of the solution tuning cycle. The graph shows a converging curve, which can be correlated with the agent's overall improvement over the iterations as it learns to avoid invalid placements, which can be attributed to $R_{Penalty}$ (refer Section \ref{sec:approach}). We observe an average decrease of 49.13\% in the aggregated number of invalid placements per workload's tuning iteration, indicating a strong positive learning trend.

\subsection{Threats to Validity}
\label{sec:threats_to_validity}

Threats to {\em Construct Validity} concern the use of controlled experimental setup and simulations. To this end, we ensured that we used workloads and configurations as close to the real-world scenario as possible. 
Specifically, we utilized the \textit{Cabspotting} dataset~\cite{Cabspotting} to train and test \approach. However, it is important to note that while our simulations provide a controlled environment for evaluation, further experiments in real-world are necessary to fully assess the practical applicability and effectiveness of our findings.

\noindent Threats to {\em Internal Validity} concern the use of a static number of nodes for placements. In practice, the DRL agent needs to be retrained if there is a change in the topology, such as the addition or removal of a node. Although we can train the DRL agent to minimize the occurrence of invalid or infeasible placements, the nature of machine learning means that we cannot guarantee that the agent will always produce valid placements. Therefore, it is essential to have an external system in place to verify the placement decisions made by the DRL agent, ensuring their correctness and feasibility.

\noindent Threats to {\em External Validity} of our approach concern the generalizability and scalability of our approach. Although our approach has been applied to two different scenarios with 4 functions and 10 functions, respectively, the techniques used in the approach are scalable to a larger number of functions. This is further validated by the results as demonstrated in Section~\ref{subsec:rq2_decision_time}. As regards to the generalizability, the approach can be integrated to any MEC system as long as it provides with mechanisms to monitor the function parameters as well as to perform routing and placement.

    

\section{Related Work}
\label{sec:related_works}
Deploying applications on edge infrastructures has increasingly become the preferred method to meet the rising demand for low-latency applications \cite{DBLP:conf/cns/VierhauserWR22}. 
Thus, the placement of such applications and their request routing in edge systems has become a primary research focus~\cite{DBLP:conf/pdp/RussoCP23} since existing solutions dedicated to cloud-computing often neglect the unique challenges of the edge context \cite{DBLP:journals/fgcs/BellendorfM20} such as managing the geographical distribution of computing nodes, maintaining low network latency, and coping with resource constrained nodes~\cite{DBLP:journals/internet/RaithND23}. Numerous studies have tackled the challenges of placement and routing in edge computing~\cite{DBLP:conf/percom/RussoMCP23,DBLP:conf/ucc/RaithRDRCR22}. A common approach involves framing the problem of service placement and workload routing as a \textit{Integer Programming} problem~\cite{DBLP:conf/icsoc/LaiHACHG018, DBLP:journals/tcc/LaiHGCAHY22}.

For example, NEPTUNE \cite{DBLP:conf/seams/BaresiHQT22} utilizes a MILP formulation to place serverless functions on edge nodes. 
Given computed placement and routing policies, NEPTUNE optimizes resources using vertical scaling controllers based on control-theory \cite{quattrocchi2024autoscaling,DBLP:conf/europar/BaresiQ18}.  Compared to NEPTUNE, \approach shares similar objectives but employs Reinforcement Learning for computing placements. This allows for timely computation of new placements and better adaptation to dynamic and fluctuating environments, such as edge computing. We view our approach as complementary to NEPTUNE: \approach can be embedded within NEPTUNE communities to determine placements, while a simplified version of NEPTUNE can then be used to compute routing policies based on these placements.

Unlike \approach, which reduces latency by placing applications closer to users, Ma et al.~\cite{DBLP:journals/tpds/MaZCHQXCWL22} focus on maximizing edge node utilization using MILP without considering network delays as we do.
Liu et al.~\cite{DBLP:journals/tsc/LiuZHXZ23} address network delays by prioritizing requests based on criticality and response times, although the approach doesn't explicitly aim to minimize delays. \approach does not explicitly consider the criticality of functions. However, one can prioritize critical application by affecting the ordering mechanism of functions employed in our RL-based approach.
Finally, Tong et al.~\cite{DBLP:conf/infocom/TongLG16} utilize Mixed Nonlinear Integer Programming to maximize served requests in a hierarchical MEC network. 
The main benefit of \approach compared to ones based on combinatorial optimization is to provide solutions in an efficient and timely manner which allow to better cope with edge nomadic users and highly-fluctuating workloads.

Raza et al. \cite{cose:23} present COSE, a framework that uses Bayesian Optimization to find the optimal resource configuration and placement for serverless applications, minimizing cost while meeting performance objectives. 
COSE provides an efficient, non-combinatorial, solution to the problem; however, compared to \approach, it focuses on cloud architectures and does not consider the intrinsic characteristics of edge topologies.

Xu et al \cite{xu:22} propose an adaptive function placement framework utilizing a Markov Decision Process to optimize serverless computing performance across terminal devices, edge nodes, and cloud data centers. The framework supports 
adaptation in real-time to allocate functions dynamically, aiming to minimize execution costs while maintaining performance satisfaction. Compared to the \approach, the approach does not consider memory requirements, network delay among edge nodes, and routing times, which are central for edge computing. Moreover, while the approach aim to optimize the utilization of the available devices, \approach also minimizes the overall network delay.

\section{Conclusions}
\label{sec:conclusions}
In this paper, we introduced \approach, an innovative method that integrates Deep Reinforcement Learning (DRL) with traditional optimization techniques, specifically Mixed Integer Linear Programming (MILP), to tackle the challenge of serverless function placement in edge infrastructures. Our evaluation shows that \approach makes decisions that are near-optimal in terms of cost and delay, while also achieving low decision-making times. For future work, we aim to leverage workload predictions to proactively place function instances and foresee potential resource saturation.

\bibliographystyle{splncs04}
\bibliography{references}
\end{document}